Structural maturation of myofilaments in engineered 3D cardiac microtissues characterized using small angle X-ray scattering

Geoffrey van Dover, Boston University, NY 14850
gdover@bu.edu     +1 (607) 793-3076

Josh Javor, Boston University, MA USA 02215, jjavor@bu.edu
Jourdan K. Ewoldt, Boston University, MA USA 02215, jewoldt@bu.edu
Mikhail Zhernenkov, Brookhaven National Laboratory, Upton NY USA 11973, zhernenkov@gmail.com
Patryk Wąsik, Brookhaven National Laboratory, Upton NY USA, 11973, pwasik@bnl.gov
Guillaume Freychet, Brookhaven National Laboratory, Upton NY 11973, gfreychet@gmail.com

Josh Lee, Boston University, MA USA 02215, hejlee@bu.edu
Dana Brown, Fort Valley State University, GA USA 31030, danabrown1621@gmail.com
Christopher S. Chen, Boston University, MA USA 02215, chencs@bu.edu
David J. Bishop, Boston University, MA USA 02215, djb1@bu.edu

**Abstract**

Understanding the structural and functional development of human-induced pluripotent stem-cell-derived cardiomyocytes is essential to engineering cardiac tissue that enables pharmaceutical testing, modeling diseases, and designing therapies. Here we use a method not commonly applied to biological materials, small angle X-ray scattering, to characterize the structural development of human-induced pluripotent stem-cell-derived cardiomyocytes within 3D engineered tissues during their preliminary stages of maturation. An X-ray scattering experimental method enables the reliable characterization of the cardiomyocyte myofilament spacing with maturation time. The myofilament lattice spacing monotonically decreases as the tissue matures from its initial post-seeding state over the span of ten days. Visualization of the spacing at a grid of positions in the tissue provides an approach to characterizing the maturation and organization of cardiomyocyte myofilaments and has the potential to help elucidate mechanisms of pathophysiology, and disease progression, thereby stimulating new biological hypotheses in stem cell engineering.

**Introduction**

The leading cause of death in the world is cardiovascular disease[1]. Research into methods for characterizing changes in cardiomyocytes during cardiac repair, regeneration, and disease progression is essential for better understanding these processes. Given the importance of the sarcomeric organization of cardiac tissue, methods for structural characterization are particularly critical to this effort. Scientists and engineers have developed optical methods[2–4], electron microscopy (EM) techniques[2,5,6], and other approaches[7–9] to characterize sarcomeric organization, but these tools for nanoscale measurement of intact tissue are limited to information from at most a few tens of microns from the surface, resulting in incomplete characterization of the subcellular structure.

Over the last decade, the use of human-induced pluripotent stem-cell-derived cardiomyocytes (hiPSC-CMs) as a model for studying the development of cell structure and function has advanced rapidly, including approaches to fabricate three-dimensional (3D) multicellular cardiac tissues, leading to improvements in cell growth, cell organization, and metrics of structural and functional maturation[10,11] (cell morphology[12], contractility[13,14], and organization of the myofibrils[12]). To improve methods of cultivation, tissue engineers have studied cardiac functions including contractility as well as cell organization as a result of confounding factors including substrate and tissue platform designs[15,16]. Despite such advances, human-induced pluripotent stem-cell-derived cardiomyocytes (hiPSC-CMs) are not yet capable of fully mimicking the structure and function of in vivo (adult) human heart tissue. At present, hiPSC-CMs generally produce beat forces tenfold smaller than adult human tissue[14]. The anisotropy of engineered cardiac tissues also shows a lower level of organization than in human heart samples (lower stiffness[9], lower contractile force[14].) This is most likely due, in part, to disorganization of the myo-actin filaments, as striations and alignment of myo-actin filaments have been shown to correlate with increased contractile force[17].

There are a limited number of approaches to characterizing the nanostructure of cardiomyocytes. Optical microscopy is a core technology that has the advantages of being relatively easy to use and inexpensive, and is capable of resolution on the scale of 0.1 µm[18]. It is ideal for the examination of cell matrices (1-100 µm), although it offers only a limited ability to resolve subcellular structures. Electron microscopy has a resolution (0.5 nm) superior to that of a light microscope. While ideal for the visualization of nanometer-scale cellular structures, EM has a depth of penetration of 1 µm, allowing only surface visualization of a multicellular complex[19]. To achieve 3D analysis with EM, samples must be fabricated using cryo-immobilization and desiccation, followed by slicing out small regions with a focused ion beam[10]. The sample is then measured in high vacuum. Bioimaging with EM has led to many discoveries such as the interacting-heads motif in the super-relaxed state[20]. Nonetheless, EM is a destructive imaging technique requiring evacuation of volatile materials essential to life and so cannot be used to image live tissue.

Compared to EM, small angle X-ray scattering (SAXS) has fewer disadvantages for studying biological tissue if radiation damage is managed[21] through a combination of microfocusing, limiting exposure time, beam intensity, and beam energy. The X-rays used in SAXS penetrate an entire cell, providing information from all the material intercepted by the beam. SAXS is capable of a higher collection rate than EM and can detect structural features in the range of 5-50 nm depending on the configuration. The spatial scale of SAXS measurements, paired with its compatibility with biological samples, enables SAXS to be unmatched for the measurement of the myofilament lattice spacing of cardiomyocytes (commonly found[21,22] to be between 30 and 50nm.)

Information about the nanostructure of cardiomyocytes has been accumulated in SAXS studies performed on cardiac tissue from humans, rats, and zebrafish[21–23]. The capabilities of SAXS for the characterization of the dynamics of the nanostructure of live cardiomyocytes were showcased by Brunello *et al.*[23]. There, the structure of cardiac tissue was studied to determine the crossbridge density of the myosin heads and the conformation of the myosin heads during contraction. The use of striated, demembranated, mature rodent tissue helped to enable a high signal-to-noise ratio for this measurement. The findings suggest that an integrative approach that incorporates subcellular insights can lead to more precise control over tissue development.

Engineered cardiac tissues do not exhibit the degree of structural maturity seen in native tissue, resulting in a lower signal-to-noise ratio in a SAXS measurement. While in recent years there has been an increased interest in the characterization of the phenotypes of hiPSC-CMs[24,25], there is still a lack of information regarding the nanoscale structure of these cardiomyocytes. Javor *et. al. (2021)* reported the first SAXS measurement of engineered cardiac microtissues, revealing that the myofilament lattice in hiPSC-CMs is a loosely ordered structural system. The spacing of the myofilament lattice at day 7 post-seeding in the cardiomyocytes was found[21] to be approximately 44 ±1.5 nm. This is distinct from the results of measurements of native adult heart tissue which found[22,26] the spacing to average 40 ±3 nm, indicating larger spacing between filaments in cardiac microtissues (CMTs) which may decrease the probability of myosin binding required for strong contraction. The resolution available in the SAXS technique, along with advances in background suppression, subtraction, and spatial averaging, make SAXS a particularly effective method to study the myofilament structure of CMTs.

Having a reliable method for measuring the structure of the cardiomyocyte may provide insights into the structural variability in addition to external factors that affect tissue growth and organization. Characterizing the nanostructure of immature, multicellular systems can provide critical information to improve cultivation methods and bridge the gap between bioengineered and natural heart tissues. The present study capitalizes on the semi-crystalline arrangement of the myofilament lattice to detect structural changes in hiPSC-CMs using SAXS. We present an approach employing a 3D tissue platform with an experimental method that facilitates extensive data collection over the entire microtissue using a 3x3 grid of measurements in conjunction with automated algorithms for data analysis. This method increases the robustness of the SAXS measurements by minimizing the impact of intrasample variation. We exploit this method to determine the myofilament spacing as a function of maturation time and as a function of position on the tissue as visualized using colormaps.

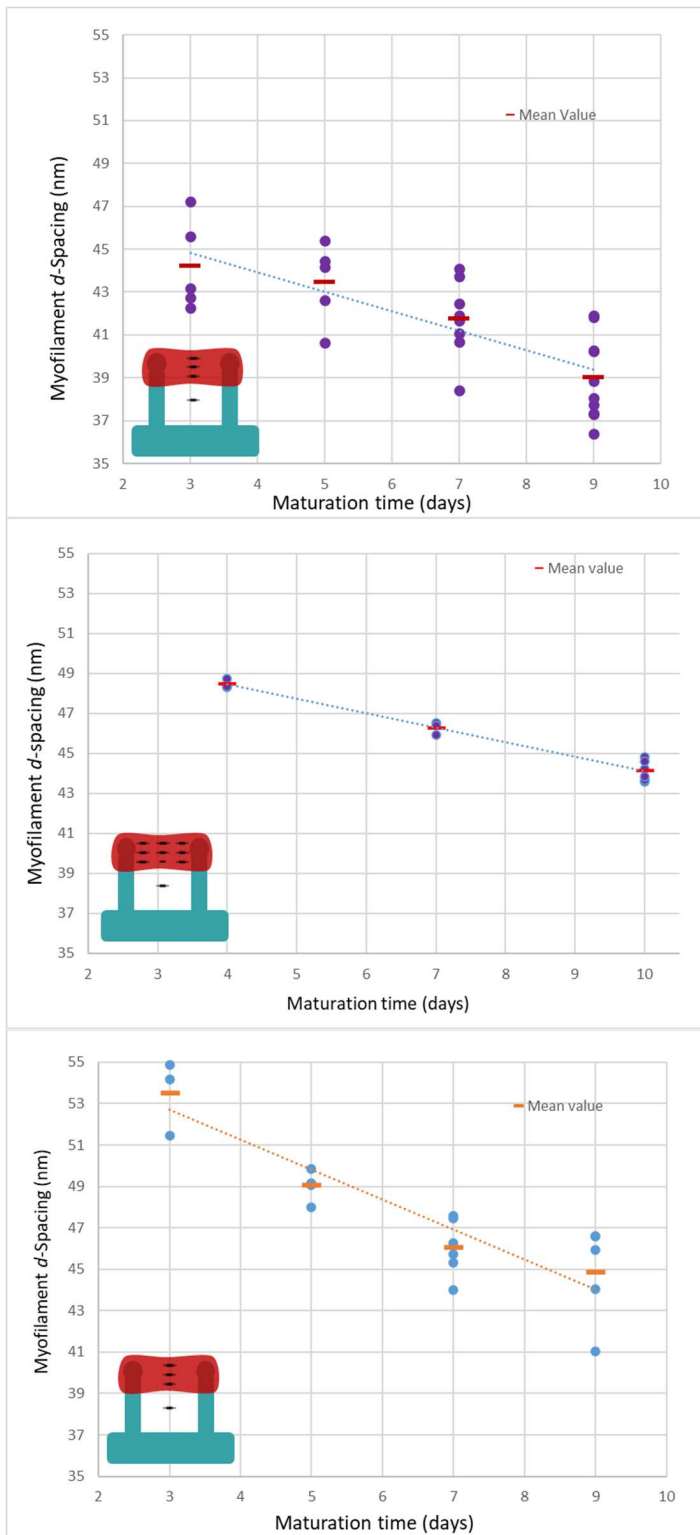

Figure 1: d-spacing as a function of maturation time, with linear regression (p<0.05). Each data point represents one entire tissue. Insets: schematic of the tissue and sample holder with black ellipses representing location of SAXS measurements; the spot not on tissue is a background measurement (SI figure 4). Top: data for Group 1 (n=28 tissue samples matured to days 3, 5, 7, and 9). Middle: data for Group 2 utilizing the 3x3 grid measurement system (n=12 matured to days 4, 7, and 10). Bottom: data for Group 3 (n=19 matured to days 3, 5, 7, and 9).

**Results**

Two batches of fixed wild-type CMT's were tested in this study. Group 1 contains wild-type tissue matured to days 3, 5, 7, and 9. Group 2 contains wild-type tissue matured to day 4, 7, and 10. Maturation time points differ between Groups 1 and 2 because of the loss of one tissue sample batch in Group 2 eliminating one data timepoint. The early time points are chosen because days 3/4 are a reliable time for the tissue to compact, and the later time points of days 9/10 were chosen as a reliable time for maintenance of tissue stability.

The primary finding of this study is that the spacing of the myofilament lattice in newly seeded hiPSC-CMs decreases over the first ten days of maturation [Figure 1]. Wild-type tissue matured to days 3-4 have a mean *d*-spacing [see Materials and Methods] of 44 nm (Group 1) and 48 nm (Group 2). The mean myofilament *d*-spacing for Day 7 samples is in the 42-46 nm range (Group 1&2), a value consistent with previous studies in myofilament spacing of hiPSC-CMs[21]. Tissue matured to days 9-10 have a mean myofilament spacing of 39 nm (Group 1) and 44 nm (Group 2). The myofilament lattice spacing is found to significantly decrease with maturation time ($p < 0.05$) for all groups. The difference in the definitive myofilament spacing between Groups 1 and 2 is due to both batch variability as well as a change in the method of measurement allowing for more robust results from Group 2. Due to the limited sample size, no definitive conclusion can be made about the average myofilament spacing at specific timepoints nor about how much the lattice spacing changes over ten days. This study focuses on the observed trend of decreasing myofilament spacing with maturation time.

Two methods of data acquisition are tested: method one employs three measurements down a column in the center of wild type tissue (Group 1), and method two employs a 3x3 grid also centered on wild type tissue (Group 2) covering between 10-20% of the tissues' cross-sectional area. The SAXS measurements are averaged to provide a single data point per tissue sample, representative of an entire tissue and between 150 (method one) to 450 cells[21] (method two). For Group 1 this means three averaged measurements and for Group 2 nine averaged measurements. The average myofilament spacing varies between batches; however, two-way ANOVA (Between Groups vs Maturation Time) results in a *p* value of 0.92, showing that the variation is statistically insignificant. By measuring the periodic spacing of the myosin filament lattice in multiple regions of the same cardiac microtissue, averaged values are obtained for a given tissue that minimize inconsistencies such as tissue deformation, intensity (SAXS beam imperfectly aligned with tissue), and inevitable sample-to-sample variance in biological materials. After averaging, using method one, the variance of the *d*-spacing between samples for Group 1 is 1.86 nm, while for Group 2 it is 0.71 nanometers. Applying method two to Group 2, the intersample variation is decreased to 0.28 nm. The reduction in the intersample variation due to averaging across a 3x3 grid allows robust conclusions even when the intersample variation is large (See SI figure 1). For experiments in which the same exact position was measured with multiple exposures, negligible variation was observed in the diffraction pattern. The major finding of this study is the trend of decreasing myofilament spacing as a function of maturation time, a finding that is found to be consistent across experiments.

To further verify the trend of decreasing myofilament spacing with maturation time, a mutant cardiomyocyte in which the myosin is hypercontractile was studied to see if contractility affected the measured d-spacing. In this sample (Group 3, single-column method, Hypertrophic Cardiomyopathic [HCM], fixed), the average myofilament spacing of HCM tissue differed from the measurements made in Javor, *et. al.* (2021) due to batch variability precluding a strong conclusion about absolute d-spacing associated with hypercontractility. However, the major finding of decreased myofilament spacing with maturation time is preserved. This finding is statistically significant ($p<0.05$). Thus, this finding is consistent across all tissue batches.

Figure 2 presents a series of interpolated two-dimensional (2D) colormaps depicting the myofilament spacing (mapped into the color scheme) as a function of position for samples from Group 2. The colormaps reveal the variation in the density (inverse of *d*-spacing) of the myosin filaments in the cardiac tissue.

It was hypothesized that the *d*-spacing would be generally lower at the extrema of the tissue where there is the most tensile stress on the cells due to the connection of the tissue to the polydimethylsiloxane (PDMS) pillars[27] (further detail in Supplementary Material figure 2 and 4). A negative gradient in the *d*-spacing is observed in the y-direction indicating some anisotropic growth a likely result of the microplatform design (SI figure 7).

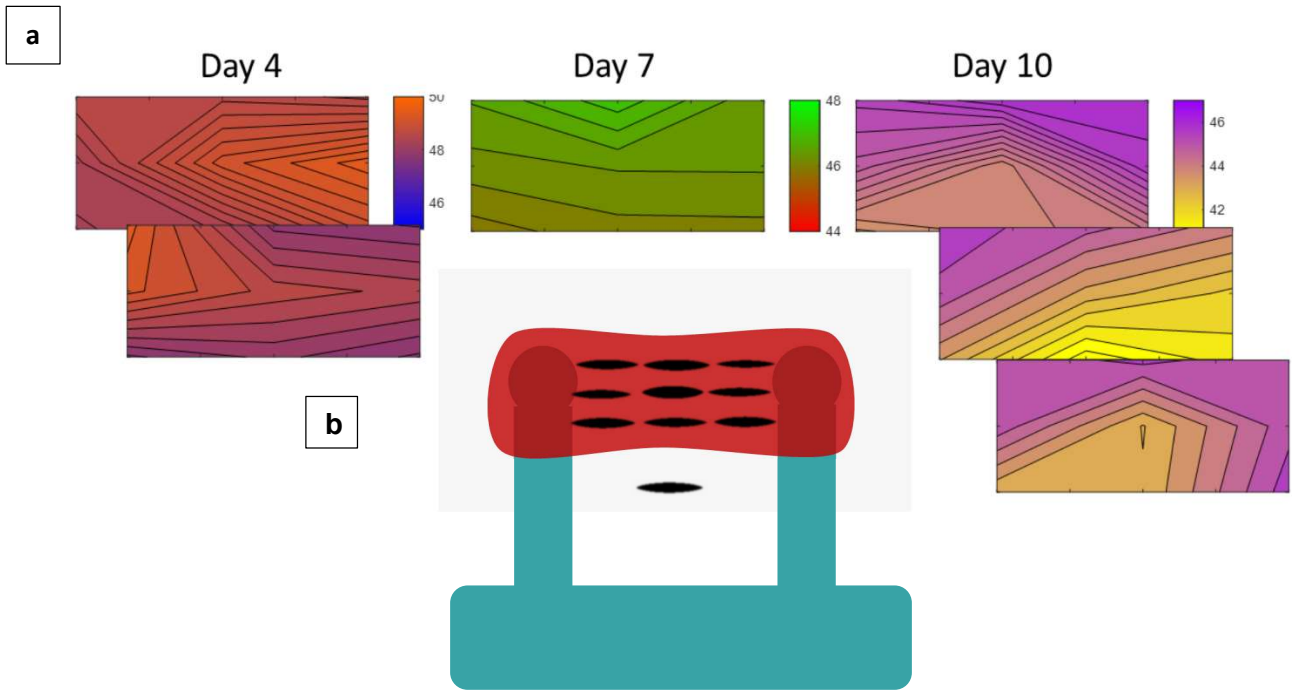

*Figure 2: (**a**) Interpolated colormaps, with calculated contour lines, of the myofilament spacing (nm) as a function of position(x&y) based on the nine positions indicated in panel b. Each column is tissues measured at specific maturation times. Extra maps are provided for days four and ten. Colorbars are adjusted for different maturation times to highlight local variation in the tissue. (**b**) visual of location of SAXS measurements on tissue as well as representation of orientation of colormaps. Note that the scale for the colormaps is different according to their maturation time. This was done so that minute changes in the myofilament spacing could be observed.*

## Discussion

The spacing of the myofilament lattice in bio-engineered hiPSC-CMTs clearly changes over the first ten days of maturation after seeding. The myofilament spacing of wild-type tissue in Group 2, for example, is shown to decrease with time, such that tissue matured for three days shows an average *d*-spacing of 48 nm and after seven days (i.e., Day 10) a decreased average *d*-spacing of 44 nm (significant at $p < 0.05$.) A lower range of d-spacing for tissue matured to days 3-4 is suggestive of insufficient time for development of cellular organization. Nevertheless, after ten days of maturation, the tissue measured in this study had an average spacing of 43 nm. Adult cardiac muscle has previously been reported[22] to be 40±3 nm. This suggests structural differences between matured adult cardiac tissue and maturing bioengineered hiPSC-CMTs. The consistency of the trend of decreased myofilament spacing with maturation time in the experimental group that utilized the 3x3 grid indicates that this structural change is occurring throughout the tissue.

Higher *d*-spacing is observed in the center of the tissue matured to Day 4 as compared to the periphery, a variation not observed in tissue matured past Day 7. There is a possibility that environmental factors can affect the myofilament *d*-spacing in the most immature tissue; if so, then this effect will be most observable at the tissue's extrema. The coordinated contraction of actomyosin fibers, coupled with the application of mechanical stress, synergistically contribute to the intricate organization and formation of sarcomeres[28]. The *lack* of variation in matured tissue implies a primacy of time-dependent cellular organization over external factors. Furthermore, a greater degree of horizontal conformity in the tissue is observed in the Day 7 samples, suggestive of a higher level of ordering (further detail in Supplementary Material figure 7). This mapping method (3x3 grid) can be used in future studies to test the effect of mechanical stress on the *d*-spacing as well as to test for anisotropy of tissue that is cultivated *in vitro*, as described in the methods section.

Definitive conclusions about systemic intrasample variation cannot be inferred from the 2D colormaps of 3D tissue due to the small sample size. The results, however, qualitatively show that there is value in collecting X-ray scattering data at a dense grid of positions in the cardiac myofilament tissue (CMTs) to observe possible spatial dependence of the subcellular structure. These 2D colormaps qualitatively demonstrate that, as the tissue matures, the myofilament spacing shows more signs of horizontal uniformity. This suggests higher cellular organization in more matured tissue. There is some indication of larger myofilament spacing at the bottom of the tissue as compared to the top. That could be due to a variety of factors including the growth patterns of the cardiomyocytes on the platform.

Many factors affect cellular structure within biological materials, but not all are currently understood and well-characterized. Since unknown confounding factors cannot be controlled effectively (i.e., batch variability, growth conditions, effect from chemical fixation etc.), this can lead to large sample-to-sample variance in structural measurements of tissue. However, while the uncertainty may be large for measurements at a specific position in a tissue, the combined measurements suggest a generalizable structure with little sample to sample variance.

The strain created by the PDMS pillars while the tissue is contracting is a possible influence on the myofilament spacing. Characterization of the effects of strain on the lattice spacing would require averaging batches grown on pillars made of materials with different compliances and is a promising direction for future studies of these tissues and their nanostructures.

The grid-style measurement employed in this study has yielded the first map of the myofilament spacing in a hiPSC-CM sample. Through this 3D characterization, changes can be monitored in the uniformity of the spacing in the cardiac tissue as it matures and can be used to determine whether cellular organization occurs universally or not. By measuring the tissue in a grid of positions and interpreting the averaged collection of measurements, conclusions about the overall structure of a tissue are robust, as compared to less reliable conclusions based on a measurement at a single localized position. Increasing the number of independent measurements taken on a given sample could further decrease the uncertainty of the myofilament structure. This work concludes that, using a 3x3 grid, the myofilament spacing of immature hiPSC-CMs has a standard deviation of 0.28 nm, a value that is significantly less than the standard deviation derived from doing only a single measurement on each sample, 2.5 nm. The HCM tissue showed similar *d*-spacing to the wild-type tissue in its development at Day 9. Chemical fixation is known to shrink the myofilament spacing by 10%, we expect that the effects of chemical fixation are independent of the degree of maturation of the tissue. The myofilament spacing values collected in this study show changes in the subcellular structure of bio-engineered cardiac tissue that are independent of tissue variability. Knowledge of the developmental differences between HCM tissue and wild-type tissue could lead to the design of new therapies to reduce the likelihood of cardiac diseases morbidity in patients.

**Conclusion**

This work takes advantage of the ability of a high brilliance synchrotron X-ray beam to form a small footprint and thereby collect the signal of a subtle aspect of the nanostructure of cardiomyocytes with an adequate signal-to-noise ratio in a transmission geometry. The myofilament spacing of hiPSC-CMs is shown to vary systematically with maturation time. The finding of decreased myofilament spacing with maturation time has a low variance and is consistent across all three experimental groups, thus, lowering the requirement for a larger sample size for the purpose of observation and statistical verification. Robust conclusions (i.e., significant decreases in *d*-spacing with tissue maturation) are made possible by improvements in cultivation, sample preparation, SAXS configuration, and sampling strategy, along with the small X-ray footprint that enables multiple measurements on a single cell to be averaged resulting in a small intersample variability. In future studies, the X-ray beam can be focused to an even smaller footprint than was utilized in this study, offering the possibility to create even more detailed spatial maps.

Our methods and results point the way for future experiments that characterize the subcellular nanostructure of hiPSC-CMs. This approach can yield insights into the relationship between mechanical stress and filament spacing, as well as characterizing and understanding the role of maturation and anisotropy on cell structure and properties. All these possibilities are relevant not only to fixed wild type tissue, but also to live and mutated tissues. Understanding the subcellular structure of matured adult human tissue and immature cardiomyocytes is the key to unlocking therapies and better systems for growing bioengineered hiPSC-CMs that can model naturally derived cardiomyocytes.

**Materials and Methods**

To cultivate and characterize hiPSC-derived cardiomyocytes, hiPSCs from the PGP1 parent line and CRISPR-Cas9 PGP1-edited cells with a heterozygous R403Q+ mutation in the β-myosin heavy chain (MYH7) are received from the Seidman Lab[29]. The cells are then cultivated using methods described by Javor, *et al.* (2021).[21] The stem cells are differentiated into cardiomyocytes through small molecule, monolayer-based manipulation of the Wnt signaling pathway[30].

A previously reported design for cardiac micro-tissue (CMT) devices with tissue wells is used for the cultivation of the CMTs[27]. These devices are equipped with two micropillars with spherical caps designed to suspend the tissue matrix. The devices are cast in polydimethylsiloxane using a 3D printed mold (Protolabs, Maple Plain, MN.) The devices are then treated using methods described in Javor *et al*. (2021).[21] The tissue is then cultivated for minimum 23 days to promote differentiation. Post seeding, each device has roughly 60,000 cells, comprising 90% hiPSC-CMs and 10% normal human ventricular cardiac fibroblasts (NHCF-V.) The cultivation process continues with methods described by Javor, *et al.* (2021).[21] In this process, Pluronic F-127 (Sigma) is added to each well to prevent cell adherence to the bottom of the well or the pillars. Tissues are maintained/matured in an incubator at 37°C and 5% $CO_2$ between 3 and 10 days.

After the chosen maturation time and after biological measurements, all the tissue is fixated using polyformaldehyde. Twenty-four hours before transportation, excess PDMS is manually removed from the devices using a scalpel with a standardized method, and the isolated tissue on its PDMS platform is transferred to a new container, washed with PBS, and placed back in the cold room.

For SAXS measurement, the tissue samples remain stretched on the pillars of the PDMS platforms and the platforms are placed in custom chambers constructed using a 70μm-thick glass slide and Kapton tape. The tissues are submerged in room temperature PBS and the Kapton/glass chamber is attached to the SAXS sample holder for diffraction studies.

Small angle X-ray scattering (SAXS) characterizes structures on the nanometer scale. X-ray diffraction can be described by Bragg's law which relates the angle of scattering of X-rays in a material to the propagation path length difference between diffracting features, thus providing an unambiguous measurement of the spacing between features[31,32]. This is the *d*-spacing of a lattice. Small features in reciprocal space relate to large scale features in real space. To characterize structures in the nanometer range, a small scattering angle is needed, requiring specialized geometry. While commonly used for crystalline samples to infer the dimensions of a lattice structure of a material, SAXS can also be used to detect semi-crystalline structures in otherwise amorphous samples[21].

**Data acquisition**

These studies employed either 14 keV (λ=0.0885 nm) (Group 1 and 3) or 12 keV (λ=0.1033 nm) (Group 2) X-rays. The 2D detector used was a Pilatus3 X 1M and the sample-to-detector distance was 7.0 m. The flux was 1 x $10^{12}$ ph/s, the sample rate was 50Hz, and exposure times of 0.09s (Group 1 and 3) or 1s (Group 2) were used. The footprint of the X-ray beam on the tissue was 20 μm x 200 μm (vertical x horizontal, respectively.) These parameters resulted in a radiation dosage of 7.6 kGy (Group 1 and 3) and 72 kGy (Group 2) (for further detail see SI eq. 1). In both cases, the dose was well under the maximum allowable dosage determined in previous studies[22]. Background data was collected by sampling from sections near the tissue, but where there is only PBS in the chamber. For multiple measurements of the same tissue, a single background sample was collected.

SAXS measurements were collected using various experimental methods. For the fixed tissue tests of Group 1, three measurements were made down a column (i.e., transverse to the long axis of the tissue) with a spacing of 100 μm, and for Group 2, nine measurements were taken on a 3x3 rectangular grid with a horizontal and vertical spacing of 100 μm (See SI figure 2). For the 3x3 grid the horizontal spacing was between each end of the X-ray beam to avoid overlap. This covers an approximate distance of 900 μm, well within the range of the length of the tissue (1 mm). The spacing between the centers of the PDMS pillars is about 1.2 mm. The 3x3 grid method was applied to only one group due to it being a recent technique and ability to corroborate the improvements was inhibited by access to the synchrotron.

The 2D diffraction pattern obtained from the Pilatus detector was azimuthally integrated using pyFAI, a Python library for fast azimuthal integration[33]. Before azimuthal integration, the image is masked and filtered, eliminating outliers resulting from dead pixels, cosmic ray strikes, blank detector regions, and other artifacts. The separately collected background is also subtracted before integration (See SI figures 3 & 5). After integration, the data is transformed into a Kratky plot, which plots $Q^2 I$ vs $Q$, to deemphasize the $1/Q^2$ dependence of a featureless signal. The Kratky plot makes it possible to discriminate the peak that reflects the *d*-spacing of the myofilaments in the tissue.

The scattering of X-rays from filaments of myosin and actin depends on the differing electron density in these filaments and provides a direct measure of the myo-actin lattice order and spacing in the volume interrogated. The SAXS scattering peak for the myofilament spacing will be approximately a wavenumber of $q = 0.145$ nm$^{-1}$, associated with a 43 nm real space lattice dimension[21,22]. To infer the position in $q$-space of the myofilament scattering peak, a linear regression was made using data where $q > 0.3$ nm$^{-1}$. Any signal from this region is due to noise or background and is not associated with the peak of interest. The linear regression was then subtracted from the signal. The myofilament scattering signal is assumed to have a normal distribution, therefore, the data is analyzed using a nonlinear fitting algorithm to determine the optimum Gaussian-function parameters. The model for the Gaussian fit provides information about the peak intensity, peak location, and full width at half maximum (FWHM.) A typical FWHM in a measurement of cardiomyocytes is about 0.01 nm$^{-1}$. The location of the Gaussian peaks indicates the wavenumber associated with the mean spacing of the cardiomyocyte in the measured area. This spacing data is collected and imported for plotting in Microsoft Excel. MATLAB is used for ANOVA analysis to test for statistical significance, and the colormap function is used to create 2D interpolated colormaps of the spacing of the myofilament lattice as a function of position on the tissue.

To make the colormaps produced in figure 2A, the nine points of data per sample collected in Group 2 measurements were formed into a 3x3 array that is then transformed into an interpolated colormap (for original data see SI figure 6). These colormaps are graphical representations of the local spacing of the myofilament spacing. Only the samples for which the X-ray beam's position with respect to the sample were known are plotted. Samples that were not well characterized (due to droop, misalignment, etc.) are not included in figure 2a. The color scheme was adjusted for each sample to display an appropriate range to highlight the variation.

**Acknowledgements**

The authors would like to acknowledge the CELL-MET program at Boston University and Beamline 12 at Brookhaven National Laboratory for valuable support.

**Conflicts of interest**

The authors declare no competing interests in relation to the work described.

**Contributions**

G.vD. planned and executed the experiments, and analyzed and interpreted the data. J. J. advised on research planning, data collection, and analysis. J. L. and J. E. cultivated the cardiomyocyte tissue. P. W., G.F., and M. Z. contributed to executing the X-ray scattering as well as processing of raw data. D. B. helped with data processing and analysis. C. C. provided expertise in the field of bioengineering, and D.J. B. provided expertise in the field of microsystems. All authors contributed to discussion of the results.

**Funding**

This research was supported by CELL-MET Engineering Research Center; NSF Award [**EEC-1647837**].

This research used the 12-ID (SMI) beamline of the National Synchrotron Light Source II, a U.S. Department of Energy (DOE) Office of Science User Facility operated for the DOE Office of Science by Brookhaven National Laboratory under Contract No. DE-SC0012704